\documentclass{ws-ijmpb}

\begin{document}

\markboth{Pavel Kornilovitch}
{Periodically Driven Adiabatic Bipolarons}

%
\catchline{}{}{}{}{}
%

\title{PERIODICALLY DRIVEN ADIABATIC BIPOLARONS}

\author{PAVEL KORNILOVITCH}

\address{Department of Physics, Oregon State University  \\
         Corvallis, Oregon 97331, USA                    \\
         pavel.kornilovich@gmail.com }

\maketitle

\begin{history}
\received{Day Month Year}
\revised{Day Month Year}
\end{history}

\begin{abstract}

Small lattice bipolarons driven by external harmonic fields are considered in the adiabatic approximation. Resonant excitation of ions modulates the trapping potential and promotes hole transfer between neighboring atomic layers. It leads to a dramatic decrease of the apparent bipolaron mass compared to the undriven case. This effect offers an explanation for dynamic stabilization of superconductivity at high temperatures recently observed in layered cuprates.   

\end{abstract}

\keywords{Bipolarons; periodic driving; high-$T_c$ superconductivity.}

\section{\label{drbip:sec:one}
Introduction
}

\begin{figure}[bt]
\centerline{\psfig{file=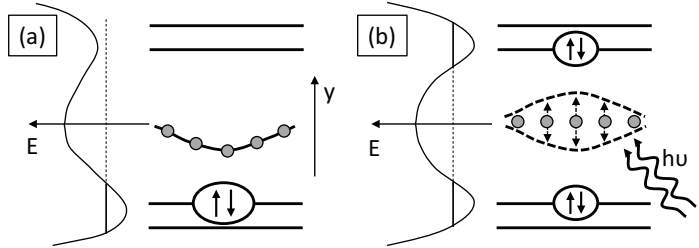,width=3.65in}}
\vspace*{8pt}
\caption{Schematic of the driven bipolaron. (a) In the absence of external fields the ions displace and stabilize a pair of holes on one bilayer. Tunneling is exponentially suppressed, as it requires simultaneous reversal of the lattice deformation. (b) An external field drives ions into resonance thereby symmetrizing the potential for holes. The tunneling probability increases exponentially.}
\label{drbip:fig:one}
\end{figure}

Recently, Hu {\em et al.} achieved stabilization of superconducting correlations in YBa$_2$Cu$_3$O$_{6.5}$ at temperatures up to 300 K by exciting apical oxygen ions with external laser fields.\cite{Hu2014,Armitage2014} The observation implies a strong coupling between ion displacements and superconducting carriers. $c$-polarized ion vibrations have been known to result in an {\em anisotropic} polaron effect with a much reduced inplane (bi)polaron mass\cite{Alexandrov1999} but much enhanced interplane mass.\cite{Kornilovitch1999} The results of Hu {\em et al.} can be easily interpreted in terms of small polaron transport between copper-oxygen bilayers. As illustrated in Fig.~\ref{drbip:fig:one}(a), in the absence of external fields, apical ions are displaced toward one of the conducting CuO$_2$ bilayers, thereby stabilizing a pair of holes in a bipolaron state. Coherent tunneling between bilayers requires simultaneous reversal of the lattice deformation. The tunneling probability $\omega_t$ is exponentially small because of small overlap of ion wave functions. In the preformed pair mechanism of superconductivity,\cite{Ogg1946,Schafroth1954,Schafroth1957,Bogoliubov1970,Micnas1990,Alexandrov1994} the Bose-Einstein condensation temperature is $T_c \propto m^{-2/3}_{ab} m^{-1/3}_{c} \propto {\omega_t}^{1/3}$. Due to the smallness of $\omega_t$, $T_c$ is exponentially suppressed. When the ions are driven into resonance by an external field, see Fig.~\ref{drbip:fig:one}(b), the trapping potential begins to oscillate symmetrically between bilayers promoting bipolaron tunneling. As a result, the tunneling frequency increases from the exponentially small $\omega_t$ to the external frequency $\omega$. It can increase $T_c$ substantially, up to several times, even after the cube root operation. This effect, possibly observed by Hu~{\em et al.}, lends support to the bipolaron mechanism of high-$T_c$ superconductivity\cite{Alexandrov1994,Alexandrov2013} that at present is not widely accepted by the scientific community. 

\section{\label{drbip:sec:two}
Adiabatic Potential
}

In this paper, {\em driven} bipolarons are investigated in the adiabatic approximation. A simple two site model is sufficient to capture the main tunneling physics. Similarly, the entire layer of apical oxygen ions is represented by a single phenomenological spring-mass quantum mode $y$. The model includes the hole part $H_{h}$ (comprising free-hole hopping and inter-hole Hubbard repulsion), the free ion part $H_{i}$, hole-ion interaction $H_{hi}$ and ion-laser interaction $H_{il}$: 
\begin{equation}
H_{h} =  
- J \sum_{\sigma} \left( c^{\dagger}_{1\sigma} c_{2\sigma} + c^{\dagger}_{2\sigma} c_{1\sigma} \right) 
+ U \sum_{s = 1,2} \hat{\rho}_{s\uparrow} \hat{\rho}_{s\downarrow}  , 
\label{drbip:eq:one} 
\end{equation}
\begin{equation}
H_{i} = - \frac{\hbar^2}{2M} \frac{\partial^2}{\partial y^2} +  \frac{1}{2} \, M \Omega^2 y^2 \: , 
\label{drbip:eq:two} 
\end{equation}
\begin{equation}
H_{hi} = - f \cdot y \sum_{\sigma} \left( \hat{\rho}_{1\sigma} - \hat{\rho}_{2\sigma} \right) \: , 
\label{drbip:eq:three} 
\end{equation}
\begin{equation}
H_{il}(y,t) = - z_i \vert e \vert \, {\cal E} \cdot y \cdot \sin{\omega t} \: . 
\label{drbip:eq:four}
\end{equation}
Here $s = 1,2$ is the site (layer) index, $\sigma = \; \uparrow, \downarrow$ is the spin index, $c$ are the fermion hole operators, $\hat{\rho} = c^{\dagger} c$, $J$ is the inter-layer single hole transfer integral, $U$ is the onsite Hubbard repulsion, $M$ and $\Omega$ are mass and frequency of the vibration mode, $f$ is the hole-ion interaction force, $z_i \vert e \vert$ is the ion electric charge, ${\cal E}$ and $\omega$ are the amplitude and frequency of an external electric field. The onsite hole energy is chosen as zero energy. Note that $U$ should be regarded as a pseudopotential. In 3D reality the bipolaron has a finite $(ab)$ size rather than being confined to a single site. In the two site model, an effective $U$ represents the real Coulomb repulsion within a bilayer. 

The above model does not include direct interaction between the external field and holes. In general, this interaction is important and leads to dynamic localization\cite{Dunlap1986} and other nontrivial effects. Here, it is neglected since dynamic localization is less prominent in a two-site system than in a long chain because the potential drop between the two sites is small. Away from special frequencies, the change in tunneling probability is expected to be small compared to the exponential increase caused by resonant excitation of ions. However, in a more complete theory yet to be developed, this interaction should be included.    

In the adiabatic approximation\cite{Holstein1959b,Kabanov1994} the hole subsystem of Eqs.~(\ref{drbip:eq:one})-(\ref{drbip:eq:four}) must be solved for an arbitrary ion configuration first. For a single ion degree of freedom, this is easily done even for {\em two} holes. Diagonalizing the $(4 \times 4)$ matrix of $H_{h} + H_{hi}$ for a fixed $y$ yields an equation for hole energy $E[y]$:
\begin{equation}
E \left\{ E \left( E - U \right)^2 - 4 J^2 \left( E - U \right) - 4 f^2 y^2 E \right\} = 0 \: . 
\label{drbip:eq:five}
\end{equation}
The lowest root of the cubic equation must be added to the ion elastic energy to form full adiabatic potential $W(y)$. For $U = 0$, 
\begin{equation}
W_{n0}(y) = \frac{1}{2} M \Omega^2 y^2 - n \sqrt{J^2 + f^2 y^2 } \: , 
\label{drbip:eq:six}
\end{equation}
with $n = 2$. $W_{n0}$ is written in this form because the choice $n = 1$ exactly reproduces the case of {\em one} hole. Thus the same expression can be used to study both the polaron and bipolaron. For $U > 0$, hole energy $E[y]$ has to be computed numerically from Eq.~(\ref{drbip:eq:five}).     

To proceed further, the adiabatic potential and the Schr\"odinger equation is transformed into a dimensionless form. Let $y_0 = \sqrt{\hbar/(M\Omega)}$ be the unit of length, $\Omega^{-1}$ the unit of time and $(\hbar \Omega)$ the unit of energy. Introducing dimensionless coordinate $\zeta = y/y_0$, time $\tau = \Omega t$, potential $w = W/(\hbar \Omega)$ and hole transfer amplitude $j = J/(\hbar \Omega)$, one obtains
\begin{equation}
i \frac{\partial \psi(\zeta,\tau)}{\partial \tau} = 
\left\{ - \frac{1}{2} \frac{\partial^2}{\partial \zeta^2} + w_{nU} - 
\alpha_{i} \zeta \sin{\left( \frac{\omega}{\Omega} \tau \right)}
\right\} \psi(\zeta,\tau) \: ,
\label{drbip:eq:seven}
\end{equation}
\begin{equation}
w_{n0}(\zeta) = \frac{1}{2} \, \zeta^2 - n \sqrt{j^2 + 2 \lambda j \cdot \zeta^2} \: .
\label{drbip:eq:eight}
\end{equation}
Equations~(\ref{drbip:eq:seven}) and (\ref{drbip:eq:eight}) contain two dimensionless coupling constants: hole-ion coupling constant $\lambda$
\begin{equation}
\lambda \equiv \frac{f^2}{2 M \Omega^2 \, J} \: . 
\label{drbip:eq:nine}
\end{equation}
and ion-laser coupling constant $\alpha_i$
\begin{equation}
\alpha_{i} \equiv \frac{z_i \vert e \vert {\cal E}}{\hbar \Omega} \sqrt\frac{\hbar}{M \Omega} \: . 
\label{drbib:eq:ten}
\end{equation}
For the system studied in Ref.~\refcite{Hu2014}, $y_0 = 5.5 \cdot 10^{-12}$ m, $\Omega^{-1} = 7.7 \cdot 10^{-15}$ s and $\alpha_{i} \approx 0.04$.         

It is useful now to summarize physical properties of the adiabatic potential, Eq.~(\ref{drbip:eq:eight}),  as it defines tunneling physics in the undriven case, $\alpha_i = 0$.

(i) The potential $w_{n0}$ becomes {\em bistable} when $\lambda$ exceeds the critical value
\begin{equation}
\lambda_{{\rm cr},n} = \frac{1}{2n} \: . 
\label{drbib:eq:eleven}
\end{equation}
The adiabatic bipolaron ($n = 2$) forms at $\lambda > \frac{1}{4}$, i.e. at a smaller $\lambda$ than the polaron (formed at $\lambda > \frac{1}{2}$). This is the essence of the bipolaron effect: two holes deform the lattice more effectively when colocated than when spatially separated.

\begin{figure}[t]
\centerline{\psfig{file=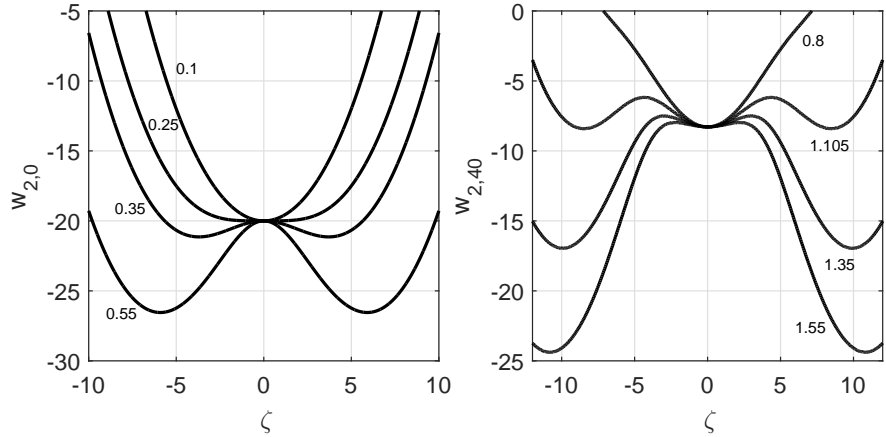,width=4.65in}}
\vspace*{8pt}
\caption{Adiabatic bipolaron potential for $J/(\hbar\Omega) = 10$. Numbers by the lines are values of $\lambda$. Left panel: $U = 0$, $w$ is given by Eq.~(\ref{drbip:eq:eight}), locations of minima by Eq.~(\ref{drbip:eq:seventeen}). Right panel: $U/(\hbar\Omega) = 40$. Note the formation of a local minimum at $\zeta = 0$.}
\label{drbip:fig:two}
\end{figure}

\begin{figure}[t]
\centerline{\psfig{file=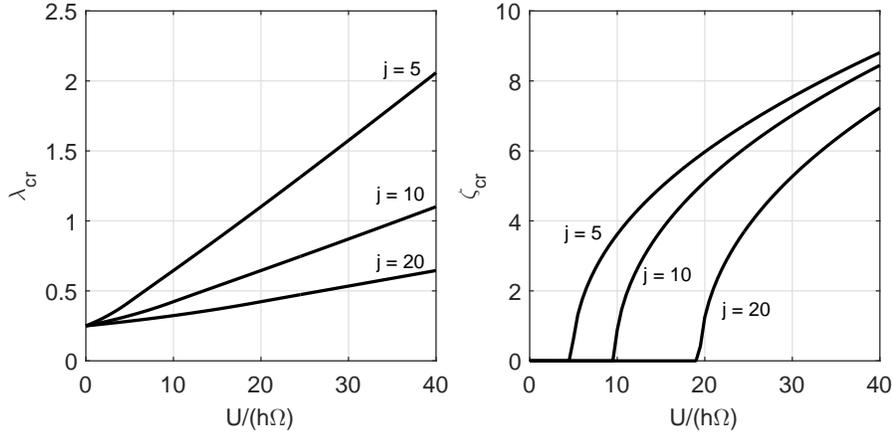,width=4.65in}}
\vspace*{8pt}
\caption{Left panel: Threshold coupling of bipolaron formation $\lambda_{\rm cr}$ for several adiabaticity parameters $j = J/(\hbar \Omega)$. Right panel: Ion displacement at bipolaron formation for several adiabaticity parameters $j = J/(\hbar \Omega)$. When the function is nonzero, bipolaron formation is akin to a first-order phase transition. The critical value $U_{\rm cr}$ at which second-order formation becomes first-order formation increases approximately linearly with $j$.}
\label{drbip:fig:aone}
\end{figure}

(ii) For arbitrary $U$, an analytical solution of the cubic equation (\ref{drbip:eq:five}) results in expressions that are too cumbersome, so it is easier to proceed numerically. Bipolaron adiabatic potential is shown in figure~\ref{drbip:fig:two} for several values of $\lambda$ and $U$. Examination of the curves reveals an interesting feature. At small $U$, the double minimum develops at $\zeta = 0$. However, at larger $U$, the new minima appear at $\zeta \neq 0$. Thus, the ground state jumps from the symmetrical point $\zeta = 0$ directly to a nonzero value $\zeta_0$. (In that sense, it is analogous to a first-order phase transition.) The zero and non-zero minima are separated by an energy barrier, see the right panel in figure~\ref{drbip:fig:two}. Figure~\ref{drbip:fig:aone} shows the coupling constant and displacement of the bipolaron potential vs. repulsion $U$, respectively. When the ion is localized in the $\zeta = 0$ minimum, the two polarons are on the opposite sides of the ion. Thus, a large Hubbard repulsion prevents bipolaron formation and stabilizes the lattice against deformation.   

(iii) Expanding $w_{n0}$ near the bottom of the well, one obtains the renormalized ion frequency 
\begin{equation}
\tilde{\Omega}_{U=0} = \Omega \sqrt{ 1 - \frac{\lambda^2_{\rm cr}}{\lambda^2} } < \Omega \: . 
\label{drbip:eq:twelve}
\end{equation}
Frequency reduction near the transition is another common feature of bipolaron systems.

(iv) At large $\lambda$, the eigenstates of the operator $-\frac{1}{2} \frac{\partial^2}{\partial \zeta} + w_{n0}$ group into pairs. The energy distance between pairs is of order $(\hbar \Omega)$ while the energy split within each pair is exponentially small reflecting the small tunneling amplitude between the wells. The ground state split can be computed analytically using the instanton technique\cite{Kabanov1994,Coleman1977,Kleinert2004} with the following result, see Appendix A 
\begin{equation}
\triangle \epsilon^{U=0}_{12} = 
\sqrt{\frac{8 j}{\pi \lambda}} \cdot F_1 \left( \frac{\lambda}{\lambda_{\rm cr}} \right) \cdot
e^{- \frac{j}{2\lambda} F_2 \left( \frac{\lambda}{\lambda_{\rm cr}} \right) } \: , 
\label{drbip:eq:thirteen}
\end{equation}
\begin{equation}
F_1(x) =  
\frac{ x^2 \left( 1 - x^{-2} \right)^{5/4} }
{\left[ x \left( 1 + \sqrt{1 - x^{-2}} \right) \right]^{ \sqrt{1 - x^{-2}} } } \: , 
\label{drbip:eq:fourteen}
\end{equation}
\begin{equation}
F_2(x) = x^2 \sqrt{1 - x^{-2}} - \log{ \left[ x \left( 1 + \sqrt{1 - x^{-2}} \right) \right] } \: .
\label{drbip:eq:fifteen}
\end{equation}
Since $F_2(x \gg 1) \rightarrow x^2$, the energy split at large $\lambda$ is $\triangle \epsilon_{12} \propto \exp{(-2 n^2 j \lambda)}$ with exponential accuracy. Thus, the bipolaron mass scales as the {\em fourth} power of the polaron mass.\cite{Alexandrov1986} For $U > 0$, the energy split can be computed numerically. With dimensionless units adopted here, the tunneling frequency between the wells is equal to the energy split: $\omega_t = \triangle \epsilon_{12}$.

\begin{figure}[t]
\centerline{\psfig{file=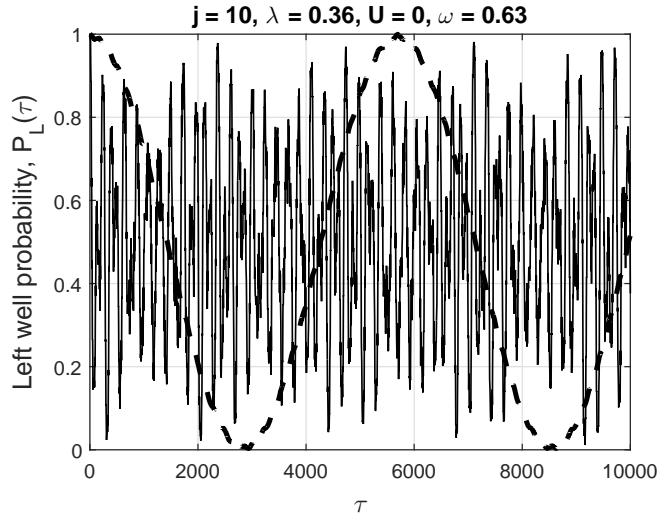,width=3.65in}}
\vspace*{8pt}
\caption{The probability to find bipolaron in the left well for $J/(\hbar\Omega) = 10$, $U = 0$ and $\lambda = 0.36$. The thick dashed is the undriven case, $\alpha_i = 0$. The thin solid line is the driven case with $\alpha_i = 0.1$ and $\omega/\Omega = 0.63$.}
\label{drbip:fig:three}
\end{figure}

\section{\label{drbip:sec:three}
Driven Bipolaron
}

The full time-dependent problem, Eq.~(\ref{drbip:eq:seven}), is now considered. The equation has been solved directly in time domain by the implicit time-stepping differencing scheme developed in Ref.~\refcite{Goldberg1967}. The numerical method is unitary and accurate to $(\triangle \tau)^2$. The method accurately reproduces the solution for a driven harmonic oscillator, which is known analytically.\cite{Husimi1953} The initial condition is a wave function localized in one of the wells 
\begin{equation}
\psi_0(\zeta) = \frac{ {\tilde{\Omega}}^{1/4}}{\pi^{1/4}} \, 
e^{- \frac{1}{2} \tilde{\Omega} \left( \zeta - \zeta_0 \right)^2 } .
\label{drbip:eq:sixteen}
\end{equation}
For $U = 0$, the well oscillation frequency $\tilde{\Omega}$ is given by Eq.~(\ref{drbip:eq:twelve}) while $\zeta_0$ follows from minimization of Eq.~(\ref{drbip:eq:eight}):
\begin{equation}
\zeta^2_{0} = \frac{2 n^2 j}{\lambda} \left( \lambda^2 - \lambda^2_{\rm cr} \right) . 
\label{drbip:eq:seventeen}
\end{equation}
For $U > 0$, both $\tilde{\Omega}$ and $\zeta_0$ are computed numerically from Eq.~(\ref{drbip:eq:five}). The quantities of interest are the probabilities to observe the bipolaron in the left (L) and right (R) wells
\begin{equation}
P_{L}(\tau) = \int^{0}_{-\infty} \hspace{-0.2cm} d\zeta \, \vert \psi(\zeta,\tau) \vert^2 , \;\;  
P_{R}(\tau) = \int^{\infty}_{0}  \hspace{-0.2cm} d\zeta \, \vert \psi(\zeta,\tau) \vert^2   \: . 
\label{drbip:eq:eighteen}
\end{equation}

\begin{figure}[t]
\centerline{\psfig{file=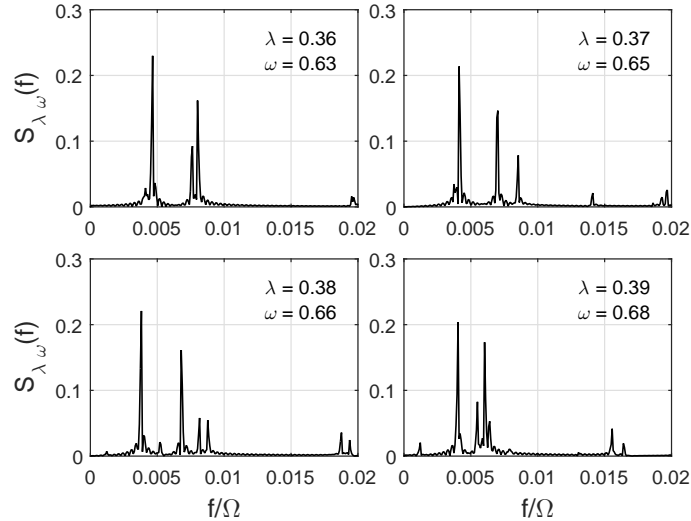,width=3.65in}}
\vspace*{8pt}
\caption{The one sided power spectrum $S_{\lambda \omega}(f)$ for $J/(\hbar\Omega) = 10$, $U = 0$, $\alpha_i = 0.1$ and $\lambda = 0.36 - 0.39$. Driving frequencies are given in units of $\Omega$.}
\label{drbip:fig:four}
\end{figure}

Figure~\ref{drbip:fig:three} compares driven and undriven $P_{L}$ for $J/(\hbar\Omega) = 10$, $U = 0$ and $\lambda = 0.36$. The energy split is $\triangle \epsilon_{12} = 0.00110$ [formula~(\ref{drbip:eq:thirteen}) is accurate to 7\%] so the undriven oscillation period between the wells is $T_0 = 2\pi/\triangle \epsilon_{12} = 5710$. Undriven tunneling between the wells is shown by the thick dashed line. Under an external field, the bipolaron begins to oscillate between the wells as shown by the thin solid line. Fourier analysis of the time series, see Fig.~\ref{drbip:fig:four} (top left), reveals a series of sharp peaks indicating coherent nature of the tunneling. The first large peak occurs at $f \approx 0.0046$, which corresponds to an oscillation period of $T = 217$. (Given a driving frequency of $\omega = 0.63$, the tunneling period is equal to about 22 driving periods.) Thus, tunneling period has been reduced by $5710/217 = 26$ times relative to the undriven case.   

Figure~\ref{drbip:fig:four} shows Fourier analysis of tunneling probabilities for several $\lambda$ and $\omega$. All cases display sharp peaks in approximately the same part of the spectrum. It indicates rough similarity in tunneling characteristics: coherent tunneling with oscillation periods between 200 and 300. This is because in all cases the external forces have close amplitudes and frequencies. The external force drives the ion into a resonance near the bottom of the well, and after about 10-12 local oscillations the ion is thrown into the other well. Since the potential is nonlinear near $\zeta = 0$, see Fig.~\ref{drbip:fig:two}, the resonant frequency changes with amplitude. Temporal dynamics becomes highly nonlinear, which generates multiple peaks in Fourier spectra. The independence of driven probabilities of the potential details is in sharp contrast with the undriven case when $T_0$ scales exponentially with $\lambda$. For example, at $\lambda = 0.39$, the energy split is already $\triangle \epsilon_{12} = 7.944 \cdot 10^{-5}$ [formula~(\ref{drbip:eq:thirteen}) is accurate to 3.7\%] and $T_0 = 79064$. The driven case still has $T = 248$, a reduction of $T_0/T = 320$ times. From that standpoint, external driving causes {\em exponential} increase of tunneling probability.

\begin{figure}[t]
\centerline{\psfig{file=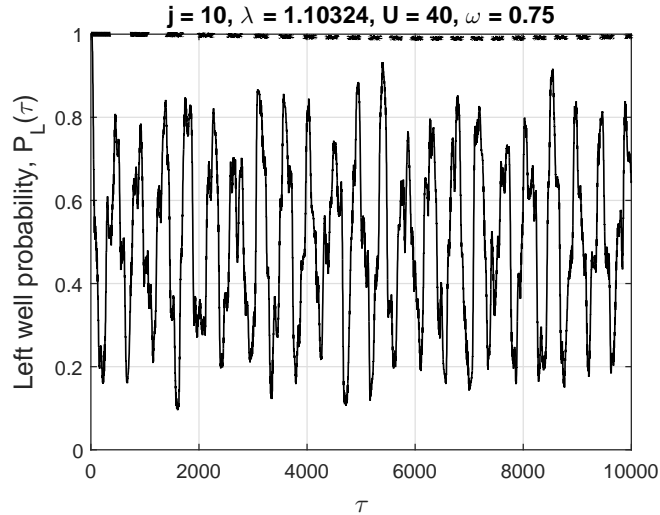,width=3.65in}}
\vspace*{8pt}
\caption{The left-well probability for $J/(\hbar\Omega) = 10$, $U = 40$ and $\lambda = 1.10324$. The dashed line is the undriven case, $\alpha_i = 0$. The solid line is the driven case with $\alpha_i = 0.1$ and $\omega/\Omega = 0.75$. }
\label{drbip:fig:five}
\end{figure}

A nonzero Hubbard potential $U$ disfavors colocation of two holes in the same layer. As a result, larger $\lambda$ are needed to form the bipolaron. As can be observed in Fig.~\ref{drbip:fig:two}, when eventually formed the double-well potential is wider than at $U = 0$ and tunneling is further impeded. The bipolaron transition is very sharp. For example, for $U = 40$, the lowest doublet energy split drops from $\approx 10^{-3}$ to $\approx 10^{-6}$ when $\lambda$ increases from 1.1032 to 1.1033, i.e. by just 0.0001. To have a reasonable tunneling without external driving $\lambda$ must be finely tuned. In contrast, under driving the tunneling probability depends on the properties of the external field and is largely insensitive to the fine details of the bipolaron potential. As an example, driven and undriven bipolarons are compared for $U = 40$ and $\lambda = 1.10324$ in Fig.~\ref{drbip:fig:five}. The oscillation period in the undriven case is $\approx 1.2 \cdot 10^{6}$ while in the driven case it is $\approx 455$, a reduction by 2700 times.     

Relevance of the obtained results to the experiments of Hu {\em et al.}\cite{Hu2014} is now discussed. Direct impact of forced oscillations on superconducting correlations is a clear evidence of strong interaction between apical oxygens and the superconducting phase in YBa$_2$Cu$_3$O$_{6.5}$. Moreover, the picture is consistent with the small polaron nature of inter-layer transport. In the absence of external forces, displaced ions trap the holes within the conducting copper-oxygen bilayers. Inter-bilayer hopping is exponentially slow because it requires simultaneous reversal of ion displacement. The process is not completely suppressed, however. The ``residual'' hopping is still sufficient to provide a macroscopic 3D coherence of the superconducting phase. The critical temperature is reduced by the factor $m^{-1/3}_c$ but nonetheless is nonzero. An external laser field drives the apical oxygens into resonance thus liberating the trapped holes. Essentially, the hole's $c$ effective mass is replaced with the external laser frequency and inter-bilayer hopping increases exponentially. As a result, superconducting weight is transferred from the intra-bilayer mode to the inter-bilayer mode, as reported in Ref.~\refcite{Hu2014}. As shown in the present work, the tunneling probability can be increased by a factor of $10^2 - 10^3$, which is sufficient to explain the 6-fold increase of the critical temperature when the cube root function is applied. Inclusion of inter-hole repulsion, represented by the Hubbard pseudopotential $U$, does not qualitatively change this picture. Rather, it shifts threshold $\lambda$ to larger values and sharpens the transition.        

The results reported here demonstrate extreme sensitivity of the superconducting properties to model parameters: bare hopping $J$, ion frequency $\omega$, hole repulsion $U$ and especially hole-ion interaction force $f$. For the superconductivity mechanism to work, the entire system must be very fine tuned. Small variations in $f$, for example, will cause exponential changes in inter-bilayer hopping probability and critical temperature. This feature may be the key to understanding why high-$T_c$ superconductivity in complex oxides is confined to a handful of compositions and is not more widespread. It can also help explain experimentally observed changes of superconductivity with chemical doping and physical pressure.

\section{\label{drbip:sec:four}
Summary
}

In summary, driven lattice bipolarons have been analyzed in the adiabatic approximation. It is shown that the bipolaron tunneling probability can be drastically increased by a periodic external force acting on the constituent ions. The ions are driven into resonance near the bottom of potential wells. The resonant conditions help accumulate energy to go over the barrier in 10 to 20 oscillations. Then the carriers follow by means of much faster bare tunneling. This results in a significant reduction of the bipolaron mass and a many-fold increase in their Bose-condensation temperature. This mechanism offers a simple explanation of dynamic stabilization of superconductivity observed in YBa$_2$Cu$_3$O$_{6.5}$. General nature of the effect suggests that similar enhancements may be observed in other layered superconductors.  Extensions of the present work beyond the adiabatic two-site case seem warranted.

\section*{Acknowledgements}

The Author wishes to thank Charles Creffield, David Roundy and Bakhrom Yavidov for useful discussions on the subject of this work, and Alexei Voronin for help with computational resources.

\appendix{\label{drbip:sec:appa}
Instanton Solution of Energy Split
}

The instanton solution of the level splitting formula\cite{Coleman1977,Kleinert2004} rests on the classical solution of the dynamic equation in imaginary time, which formally corresponds to inverting the sign of the potential $w_{0n}$
\begin{equation}
\frac{1}{2} \left( \frac{d\zeta}{d\tau} \right)^2 
- \frac{1}{2} \, \zeta^2  + n \sqrt{ j^2 + 2 \lambda j \cdot \zeta^2 } 
= \varepsilon_0  \: .
\label{drbip:eq:aone}
\end{equation}
The integration constant $\varepsilon_0$ is the energy of the double peak of the inverted potential:
\begin{equation}
\varepsilon_0 = \frac{j}{4\lambda} \left( 1 + \frac{\lambda^2}{\lambda^2_{\rm cr}} \right) \: .
\label{drbip:eq:atwo}
\end{equation}
Integrating Eq.~(\ref{drbip:eq:aone}), one obtains an instanton solution that passes $\zeta = 0$ at $\tau = 0$ with a positive velocity:    
\begin{eqnarray}
\tau[z(\zeta)] & = & \int^{z(\zeta)}_{j} \frac{z \: dz}{(2n \lambda j - z) \sqrt{z^2 - j^2}} 
\nonumber \\
               & = & \log{\frac{j}{z + \sqrt{z^2 - j^2}}} 
\nonumber \\               
               &   & + \frac{z_m}{\sqrt{z^2_m - j^2}}
\log{\frac{( z_m z - j^2 ) + \sqrt{(z^2_m - j^2)(z^2 - j^2)}}{j (z_m - z)}} \: , 
\label{drbip:eq:athree}
\end{eqnarray}
where $z(\zeta) = \sqrt{ j^2 + 2 \lambda j \cdot \zeta^2 }$ and $z_m = z(\zeta_0) = 2 n \lambda j = j ( \lambda/\lambda_{\rm cr} )$. In the limit $z \rightarrow z_m$, (or $\zeta \rightarrow \zeta_0$), this expression diverges logarithmically and can be inverted to give the large-time asymptote of the instanton:
\begin{equation}
\zeta_0 - \zeta  =  \sqrt{\frac{2 j}{\lambda}} \cdot  
\frac{ \beta \left( \frac{\lambda}{\lambda_{\rm cr}} \right)^2 }
     { \left[ \left( \frac{\lambda}{\lambda_{\rm cr}} \right) ( 1 + \beta ) \right]^{\beta}}
  \cdot e^{ - \beta \tau } \equiv C \cdot e^{ - \beta \tau } , 
\label{drbip:eq:afour}
\end{equation}
where $\beta$ is the frequency renormalization factor appeared in Eq.~(\ref{drbip:eq:twelve})
\begin{equation}
\beta = \sqrt{ 1 - \frac{\lambda^2_{\rm cr}}{\lambda^2} }   \: . 
\label{drbip:eq:afive}
\end{equation}
Thus, approach to the top is governed by potential curvature near the top, as expected on physical grounds. The prefactor $C$ plays an important role in tunneling. According to the dilute instanton gas approximation,\cite{Coleman1977,Kleinert2004} the energy split of the lowest level pair is given by  
\begin{equation}
\triangle E_{12} = 2\hbar \sqrt{\frac{\triangle A_{\rm cl}}{2 \pi \hbar}} \cdot 
\beta^{3/2} \sqrt{\frac{2\hbar}{\triangle A_{\rm cl}}} \cdot (\Omega C) \cdot \: 
e^{- \frac{1}{\hbar} \triangle A_{\rm cl} } = 
(\hbar \Omega) \cdot \frac{2 \beta^{3/2} C}{\sqrt{\pi}} \cdot
e^{- \frac{1}{\hbar} \triangle A_{\rm cl} }\: . 
\label{drpol:eq:fiftyfour}
\end{equation}
Here $\triangle A_{\rm cl}$ is the difference in classical actions between the instanton and ``in the well'' classical solutions, which is given by 
\begin{eqnarray}
\frac{1}{\hbar} \triangle A_{\rm cl}  & = & \int^{\zeta_0}_{-\zeta_0} d\zeta 
\sqrt{ 2 j \left( n^2 \lambda + \frac{1}{4\lambda} \right) + \zeta^2 - 
2 n \sqrt{j^2 + 2 \lambda j \zeta^2 }} 
\nonumber \\
                             & = & \frac{j}{2 \lambda} \left\{ 
 \frac{\lambda \sqrt{\lambda^2 - \lambda^2_{\rm cr}} }{\lambda^2_{\rm cr}} - 
 \log{ \frac{\lambda + \sqrt{\lambda^2 - \lambda^2_{\rm cr}}}{\lambda_{\rm cr}} } \right\} \: . 
\label{drpol:eq:fiftyfive}
\end{eqnarray}
Substitution of Eq.~(\ref{drpol:eq:fiftyfive}) and $C$ from Eq.~(\ref{drbip:eq:afour}) in the level-splitting formula, Eq.~(\ref{drpol:eq:fiftyfour}), leads to the final result, Eqs.~(\ref{drbip:eq:thirteen})-(\ref{drbip:eq:fifteen}). For the adiabatic polaron, it was first reported by Kabanov in Ref. \refcite{Kabanov1994}.

\end{document}